\documentclass[10pt,twocolumn,letterpaper]{article}

\usepackage{booktabs}
\usepackage{array}
\usepackage{multicol}
\usepackage{graphicx}
\usepackage{xcolor}
\usepackage[letterpaper,margin=0.62in,columnsep=0.22in]{geometry}
\usepackage{tikz}
\usetikzlibrary{arrows.meta,positioning,fit,calc}
\usepackage[hidelinks]{hyperref}
\usepackage[capitalise,noabbrev]{cleveref}
\usepackage{comment}

\definecolor{mnavy}{HTML}{132D52}
\definecolor{mblue}{HTML}{2464A8}
\definecolor{mcyan}{HTML}{12A8C7}
\definecolor{mmint}{HTML}{36AD8A}
\definecolor{mamber}{HTML}{EAA133}
\definecolor{mslate}{HTML}{5C6B7A}
\definecolor{mpale}{HTML}{F3F6F9}

\title{MimicSat: A Reconfigurable Cyber-Physical Testbed
For Small Satellite Systems and Cybersecurity Research}

\author{Nisha Vinayaga-Sureshkanth\thanks{Equal contribution}, A H M Nazmus Sakib\footnotemark[1], Mahsin Bin Akram\footnotemark[1],\\
David R. Silva and Murtuza Jadliwala \\[2pt]
\small The University of Texas at San Antonio, San Antonio, Texas, USA}
\date{}

\newcommand{\keywords}[1]{\par\noindent\textbf{Keywords:} #1\par}

\begin{document}
\maketitle
\begin{abstract}

MimicSat provides a common experimental environment for examining how changes in satellite subsystem behavior propagate to mission outcomes across software-based and hardware-based execution. Its design is motivated by controlled spacecraft cybersecurity studies involving attacks, faults, and defensive responses.
In MimicSat, a \emph{mission} encompasses the spacecraft and ground activities required to achieve defined objectives, and an \emph{experiment} consists of one or more mission runs used to study selected conditions or interventions.
To support such studies, MimicSat offers software-based and hardware-based execution environments that implement the same mission functions and data exchanges, while allowing specific functions to be realized differently. A shared mission definition preserves command and telemetry semantics across environments, and collected observations retain provenance about the originating participants and acquisition paths. As a result, mission behavior can be compared across execution configurations without redefining the surrounding mission.
This paper presents the architectural principles of MimicSat, its software and hardware execution forms, and their integrated operation. MimicSat also supports satellite systems engineering, mission operations, resilience studies, and related experimental use cases.
\end{abstract}

\section{Introduction}

Satellite behavior emerges from interactions among onboard computation, power, sensing, communications, payload activity, and ground operations. A payload acquisition may complete successfully onboard yet still fail to produce a usable ground product if storage is exhausted, power becomes unavailable, or a communication contact ends before the data can be downlinked. Similarly, a delayed command may arrive after the mission has entered a state in which the requested operation is no longer valid. These dependencies must be examined before launch because correcting an interface, firmware, or operational failure after deployment can be difficult or impossible. Therefore, preflight testing must reproduce the mission relationships that can influence an outcome while providing sufficient visibility to determine why an expected result did not occur. 

Software-based environments and physical test facilities address different aspects of this problem. The NASA Operational Simulator for Small Satellites, NOS3, supports spacecraft development and mission testing before all hardware is available \cite{grubb2016,lucas2023}. Physical and hybrid facilities extend this capability by introducing embedded execution and electrical behavior that can affect mission outcomes \cite{kiesbye2019}. Such environments become especially important when an investigation depends on firmware behavior, bus contention, finite queues, resource consumption, or recovery following a device reset. The appropriate execution configuration therefore depends on which physical and computational effects must be represented for the research question.

Comparing results across these execution configurations introduces another challenge. Software-based and embedded participants may perform the same mission function while relying on different interfaces, timing behavior, and observation mechanisms. If commands, telemetry interpretation, operating conditions, or other elements of the mission scenario also change between runs, an apparent implementation difference may instead result from those changed conditions. At the same time, preserving identical field names does not guarantee equivalent behavior because units, update rates, saturation limits, and error handling may still differ. Therefore, meaningful comparison requires stable functional definitions together with an explicit record of what changes between mission runs.

MimicSat addresses this challenge by providing independently operable hardware- and software-based execution environments that represent the same space mission. Mission roles can be assigned to embedded participants or virtual services while the surrounding mission maintains a consistent command and telemetry vocabulary. Observations remain attributable to the participant and acquisition path that produced them, allowing subsystem-reported state to be compared with independently captured communication and execution records and clarifying how local behavior contributes to mission outcomes.

Controlled spacecraft cybersecurity research is the primary application considered in this work, particularly studies that relate attacks, faults, and defensive responses to subsystem behavior and mission-level consequences. The shared mission definition and independently observable execution environments also support spacecraft engineering, mission operations, education, and resilience studies. The following sections define the experimental boundary of MimicSat, describe its architecture and execution environments, compare it with established satellite system test environments, and discuss representative research use cases.
\section{Foundations and Scope}
\label{sec:found}

\subsection{Mission and experiment}

In MimicSat, a \textbf{space mission} is represented as a coordinated set of spacecraft and ground activities carried out to achieve a defined mission objective. A \textbf{mission scenario} describes the conditions in which the mission is executed, including initial conditions, parameter values, communication conditions, and, when applicable, controlled interventions. A \textbf{mission run} is a single execution of a mission scenario under a specified configuration. An \textbf{experiment} groups one or more mission runs to address a research question or to compare selected conditions. In this way, the mission describes what the represented spacecraft and ground system aim to accomplish, while the experiment describes how mission runs are used to support investigation.
The represented mission is kept conceptually separate from the infrastructure used to observe and control the experiment. Commands, subsystem state, payload products, and telemetry are part of the mission representation. Independent acquisition, experiment controls, and protection mechanisms are part of the test environment. This separation allows mission behavior to be observed without making the observation infrastructure part of the represented flight architecture.

\subsection{Roles and realizations}
A \textbf{mission role} defines a required function, such as command handling, power management, or payload acquisition. A \textbf{realization} specifies how that function is executed, whether on embedded hardware, through a virtual service, by replay, or through an external environment. A \textbf{participant} is the specific node, service, model, or gateway assigned to that role during a mission run. The role therefore preserves the expected mission function, while the participant identity records which implementation produced a particular response.

MimicSat captures selected satellite data flows, bus behavior, and subsystem interactions and relates them to ground services and mission products. Its initial functional organization follows CubeSat-class systems. The \textbf{hardware mimic} is an execution environment in which mission roles are realized using commercial off-the-shelf (COTS) hardware. It is not intended to replicate flight electronics, spacecraft construction, or the physical design of individual subsystems. The corresponding \textbf{software mimic} is an execution environment in which the same mission roles and exchanges are realized through software services.

\subsection{Observation and evidence}

Command and telemetry are mission records, while evidence is the broader set of records used to interpret mission execution. A \textbf{command} records an intended action, its destination, and the outcome at each available processing stage. \textbf{Telemetry} reports the state a mission participant makes available to the surrounding mission or ground system. \textbf{Evidence} includes these records together with independent observations of communication interfaces, execution events, resource behavior, and experimental interventions. A received acknowledgment may show that a command reached a given processing stage, but it does not confirm that the requested mission activity completed successfully. Similarly, an observed bus transaction and a decoded telemetry value capture different points along the same information path. Preserving these distinctions makes it possible to identify where an expected mission outcome first diverged from the recorded execution.

\subsection{Fidelity and intended use}

NASA STD 7009B \cite{nasa7009} evaluates the credibility of models and simulations in relation to their intended use, verification, validation, uncertainty, and operating limits. Following this approach, MimicSat defines fidelity according to the mission behavior that must be represented for a particular research question rather than according to a single measure of physical similarity to flight hardware. The required fidelity is therefore established by the research question and reflected in the conditions represented within each mission run, including command and telemetry semantics, interaction ordering, communication behavior, timing effects, resource constraints, and observable mission consequences.

A mission role realized in hardware or software can provide different but complementary forms of fidelity. A physical bus can expose contention, device timing, error handling, and recovery behavior even when the connected nodes do not reproduce flight electronics. A software realization can preserve the required mission functions when those physical effects are not relevant to the research question.
For each mission run, the resolved configuration must identify the behaviors being represented, the realizations used to produce them, the observations available for analysis, and the assumptions that limit interpretation. Fidelity in MimicSat is consequently evaluated relative to the intended use of a configured mission run, not as an inherent property of a single realization.

\subsection{Testbed Requirements}
The preceding definitions establish MimicSat as a configurable environment in which mission functions can be assigned to different realizations and examined across mission runs. Supporting controlled spacecraft cybersecurity experiments requires more than reproducing individual subsystem functions. The testbed must also preserve the mission context, configuration, observations, intervention conditions, and recovery state needed to relate local behavior to mission-level consequences. These requirements also support related engineering, resilience, and operational studies. Therefore, MimicSat adopts the following design requirements as the basis for the architecture described in the following section.

\textbf{Mission-context preservation:} The testbed must represent the mission roles, subsystem dependencies, operating modes, commands, telemetry, communication paths, and ground interactions required to relate local subsystem behavior to mission-level consequences.

\textbf{Repeatable and explicit configuration:} Each mission run must begin from a resolved configuration that identifies the participants and realizations assigned to mission roles, initial state, timing policy, resource assumptions, communication conditions, and enabled observation channels. Repeated mission runs must reproduce these declared conditions or record every intentional difference.

\textbf{Controlled intervention:} The testbed must support bounded and authorized changes to commands, messages, sensor inputs, communication conditions, resource limits, participant availability, or operator information. Each intervention must retain its source, target, activation condition, duration, and release condition. This capability supports controlled cybersecurity experiments as well as fault and resilience studies.

\textbf{Independent and attributable observation:} Mission telemetry must be distinguishable from independently collected bus records, execution events, resource measurements, and intervention records. Every observation must identify its source and acquisition path so that participant-reported state is not treated as the sole account of system behavior.

\textbf{Cross-realization comparability:} A mission role must be transferable between compatible software-based and hardware-based participants without redefining the surrounding mission. Command meanings, telemetry fields, units, timing assumptions, interface transformations, and error behavior must be explicit so that implementation-specific effects can be distinguished from changes to the mission scenario or run configuration.

\textbf{Safe containment and recovery:} Experimental controls must be separated from ordinary mission behavior and must not override physical protection mechanisms. Each mission run must define how an intervention or fault condition is stopped, how affected participants are reset or isolated, and how evidence collected before and after recovery remains associated with the correct configuration.

\section{The MimicSat Architecture}

\begin{figure*}[]
\centering
\includegraphics[width=1\textwidth]{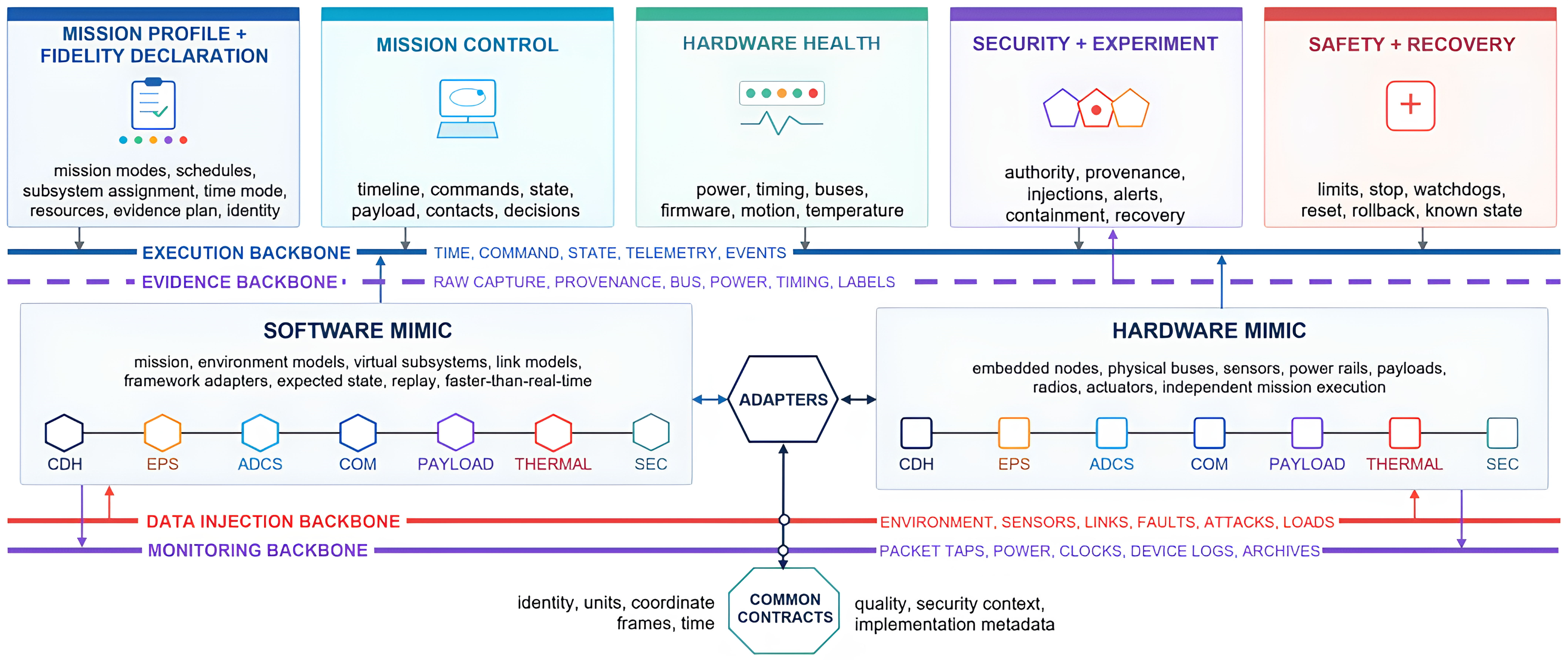}
\caption{MimicSat functional architecture. Software-based and hardware-based labels identify alternative execution realizations of represented mission roles and data flows rather than replicas of physical flight subsystems. CDH denotes command and data handling, ADCS attitude determination and control, TCS thermal control, and EPS electrical power. Integration, intervention, comparison, and monitoring indicate logical relationships rather than implementation procedures.}
\label{fig:architecture}
\end{figure*}

The MimicSat architecture (visualized in \cref{fig:architecture}) connects three responsibilities that must remain coherent across mission runs. Mission configuration defines the represented mission scenario and identifies which participants perform its required roles. Execution and observation distinguish the information that drives mission behavior from the evidence used to interpret that behavior. Intervention and monitoring introduce controlled changes and relate their consequences to time, source identity, and protection state. Together, these responsibilities address the testbed requirements defined in \cref{sec:found}  and establish the functional boundary shared by the hardware mimic and software mimic described below.

\subsection{Design Principles}

\subsubsection{Mission configuration and contracts}
A \textbf{mission profile} specifies the roles, modes, schedules, command and telemetry meanings, resource assumptions, timing policy, and observation channels required to execute a mission scenario. Resolving the profile assigns an identifiable participant and realization to each required role. The resulting configuration therefore records how the mission scenario will be executed for a particular mission run. The same mission scenario can consequently be executed using software-based, hardware-based, or mixed configurations while retaining an explicit record of the realization and participant assigned to each role.
Role assignment is kept separate from authority over operational mission state because multiple sources may represent the same mission function. When this occurs, the resolved configuration identifies which source governs mission behavior and which sources are retained for observation or comparison.

Common \textbf{mission contracts} preserve the meaning of information exchanged across participants and realizations. They define the information and behavioral expectations required for compatible mission roles while allowing participant-specific interfaces to remain implementation dependent. The resolved configuration retains sufficient information to identify the participant and realization associated with each role and observation. \cref{fig:architecture} summarizes these relationships and shows how participant assignments can change without changing the functions required by the represented mission.

\subsubsection{Execution and observation}
After mission configuration establishes the roles, participants, and contracts for a mission run, the \textbf{execution path} carries the information required for the mission to proceed. This information includes commands, acknowledgments, telemetry, events, environmental inputs, resource state, and payload products. Participant-specific interfaces conform to the common mission definitions required for compatible execution.
Successful delivery across an interface does not establish that the receiving participant interpreted the information correctly or completed the intended operation. Therefore, MimicSat retains observations via the \textbf{evidence path} that allow reported mission state to be interpreted alongside independently observed execution behavior. These observations remain associated with their sources so that disagreements between participant-reported state and externally observed behavior can be identified.
A participant may report that a payload product was transmitted while an independent bus observation records an incomplete exchange. Preserving both records distinguishes participant-reported state from behavior observed by the test environment. When evidence is missing, that absence is recorded as an explicit limitation rather than being inferred from mission telemetry.

\subsubsection{Intervention and monitoring}
Controlled \textbf{intervention} uses the configuration and evidence mechanisms established in the preceding subsections to introduce bounded changes during a mission run. These changes may affect selected inputs, messages, resource conditions, participant availability, or communication paths. Each intervention retains sufficient context to identify what was changed, where the change was applied, and when it affected the mission run. Original and altered values remain linked when both are available. Therefore, subsequent observations can relate the intervention to changes in subsystem exchanges and mission state without treating it as an unexplained configuration difference.

\textbf{Monitoring} relates command, telemetry, evidence, and intervention records using their temporal context and participant identities. An ordered display can support mission operation, but it cannot establish event order beyond the precision of the available clocks and acquisition paths. Timing uncertainty, unavailable channels, and missing records must therefore remain part of the interpretation. Physical limits, independent protection mechanisms, and de-energization controls continue to constrain each mission run regardless of the intervention.

\subsection{Hardware and Software Mimic}

\subsubsection{Distributed hardware execution}
The hardware mimic uses embedded participants assigned to specific mission roles for a mission run. Each participant processes commands relevant to its role, maintains local state, and exchanges telemetry and events over the represented data buses. Command and data handling (CDH) coordinates mission modes and routes information across participants. As a result, the hardware mimic provides distributed execution of represented mission functions without attempting to reproduce the physical construction of flight subsystems.

Communication remains part of the represented mission, instead of serving only as a connection between independent components. A command may reach an embedded participant without producing the intended activity, and telemetry may be generated locally without reaching the ground service. By preserving these intermediate states, interface behavior can be related to subsequent mission consequences. This approach allows the hardware mimic to execute the required mission roles with selected supporting inputs and ground functions, without relying on the software mimic to control those roles.

\subsubsection{Subsystem dependencies}
Subsystem roles preserve the information relationships required for mission execution. Power state can constrain permitted activity, communication state can determine whether telemetry reaches the ground, and payload state can affect storage and subsequent data transfer. These dependencies are represented through information exchanged among participants so that a change within one subsystem can produce observable consequences elsewhere in the mission.
Sensor data and environmental inputs provide context for selected subsystem functions, while reference state, information delivered to subsystem logic, and reported output retain distinct meanings. A value used for reference or comparison is therefore distinguishable from the value available to a participant. Attitude, thermal, and power roles can consequently represent mission-relevant relationships without requiring replication of corresponding flight hardware.
Similarly, a replaceable payload preserves the relationships among acquisition, processing, storage, and downlink even when its data source changes. Acquisition, onboard availability, transfer, and ground receipt can therefore be examined as distinct stages of mission behavior. The relevant correspondence lies in the behavior and consequences of these data flows within the mission rather than in reproducing the physical construction of a particular instrument.

\subsubsection{Virtual mission services}
The software mimic provides virtual participants for the mission roles and supporting functions required by a mission scenario. These participants preserve the same mission definitions used by the hardware mimic while representing subsystem, environmental, communication, payload, and ground interactions through software. Mission-relevant state and exchanged information retain their defined meanings so that software-based execution remains comparable with other realizations.
The software mimic also supports controlled variation of mission conditions and repeatable mission runs. Communication availability, timing effects, subsystem state, and other configured conditions can therefore be varied while preserving the surrounding mission scenario. Recorded mission runs may also be reused for subsequent analysis or replay, allowing previously observed behavior to be examined without treating replayed activity as the original execution.

\begin{figure*}[]
\centering
\includegraphics[width=0.98\linewidth]{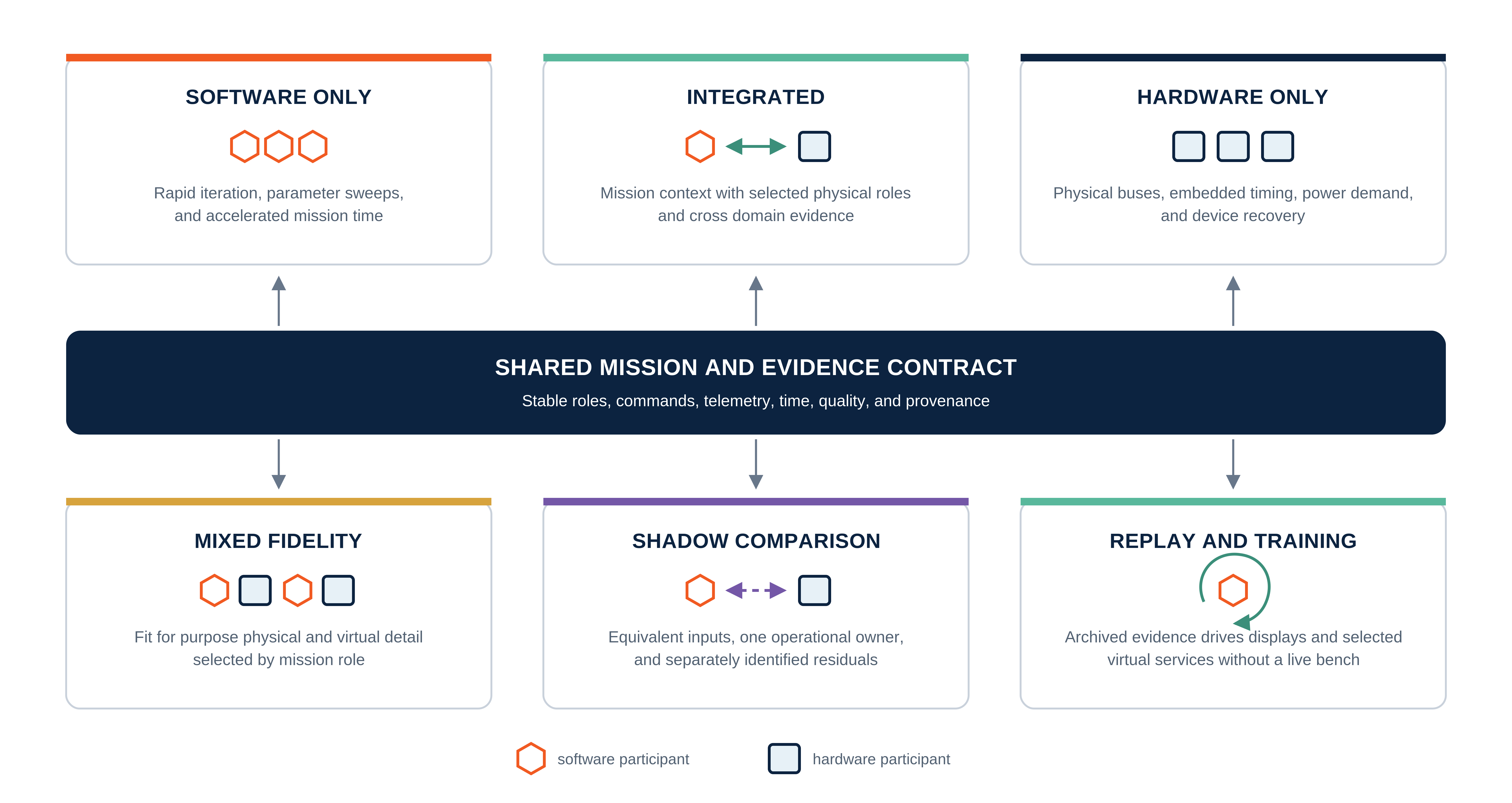}
\caption{MimicSat execution configurations. Hardware-based, software-based, and mixed configurations preserve the same mission definition while allowing mission roles to use different realizations. Shadow comparison and replay support comparison and reuse of recorded execution without redefining the represented mission.}
\label{fig:modes}
\end{figure*}

\subsection{Integrated Operations}
\subsubsection{Assignments and adapters}
Integrated operation allows selected mission roles to use hardware-based participants while other roles are provided through software-based realizations. The mission scenario does not require corresponding hardware and software instances of every role. Instead, each mission run uses the combination of realizations required by its specified configuration while preserving the functional relationships among mission roles. \cref{fig:modes} summarizes the execution configurations supported by this organization.
\textbf{Adapters} allow participants with different local interfaces to conform to the common mission representation. The configuration identifies the participant and realization assigned to each role so that implementation changes remain explicit when mission runs are compared. Differences introduced by a selected realization or its interface remain part of the experimental context rather than being treated as changes to the surrounding mission.

\subsubsection{Shadow comparison}

Shadow comparison allows an operational participant and a reference participant to receive equivalent inputs while preserving their outputs as separate observations. Only the operational participant controls mission state. Differences in reported state, timing, events, or communication can therefore be examined without allowing the reference participant to alter the executing mission.
Observed differences do not by themselves establish a fault or a difference in fidelity. They may instead result from modeling assumptions, interface transformations, timing behavior, or implementation choices. Shadow comparison therefore provides attributable evidence for analysis rather than an automatic classification of the observed difference.

\subsubsection{Recovery and continuity}

Recovery is treated as part of the mission run rather than as a transparent continuation of prior execution. Restored communication does not necessarily imply restoration of the preceding mission state. Recovery events therefore remain associated with the configuration and observations that precede and follow them.
If the participant assigned to a mission role changes during a run, the transition remains explicit so that subsequent state and observations can be associated with the appropriate realization. This preserves continuity of interpretation when execution resumes after interruption, replacement, or recovery.

\subsection{Architectural contribution}

The central architectural contribution of MimicSat is the preservation of a common mission context as individual mission roles change realization and experimental conditions are varied. Mission roles can be assigned to software-based or hardware-based participants without redefining the surrounding mission, while observations remain attributable to the participants and acquisition paths that produced them. This organization allows mission runs to be compared when implementations or interventions change, while keeping those changes distinct from changes to the mission scenario itself.

\section{Related Work}
\label{sec:related}

\subsection{Engineering and mission environments}
FlatSat engineering testbeds \cite{barcellos2023flatsat} provide physical access to spacecraft electronics, interfaces, and subsystem interactions during integration and testing. The European Space Agency FlatSat demonstration \cite{esa2021} uses an opened-out CubeSat whose onboard computer, communication system, attitude control, power system, navigation receiver, and other hardware remain accessible on a laboratory bench. The platform supports end-to-end operation through a ground station, including radio commands and telemetry downlink. Such facilities allow integrated spacecraft behavior to be examined while retaining direct access to physical subsystems.

Software environments provide a complementary capability by supporting mission development before all physical hardware is available. NOS3 \cite{grubb2016,lucas2023} integrates flight software, spacecraft hardware models, dynamics simulation, and ground interaction within a common development environment. It also supports message monitoring, interception, injection, and combinations of simulated and physical components. The NOS3 development workflow \cite{nos3validation} supports progression from software simulation toward testing with spacecraft hardware.

MimicSat addresses a different experimental objective in which its hardware mimic and software mimic provide independently operable execution environments for the same mission, while allowing individual mission roles to use different realizations. Mission roles, command meanings, telemetry semantics, and expected interactions remain consistent across execution configurations. Selected mission functions can therefore change realization without requiring the surrounding mission to be redefined. This distinction is important when the research objective is to determine whether observed differences arise from implementation behavior, such as timing, firmware, communication, or resource effects, rather than from changes to the mission scenario itself.

\subsection{Hybrid infrastructure}
Hybrid test environments introduce physical effects that software models may not fully reproduce. MOVE II \cite{kiesbye2019} combines software-in-the-loop and hardware-in-the-loop testing with simulated environmental interactions and supports attitude control testing, electrical power analysis, operator training, and investigation of flight anomalies. The JPL Small Satellite Dynamics Testbed \cite{sternberg2018,rivera2023} similarly combines simulation with physical experimentation for guidance, navigation, and control research. FlatHILS \cite{greenberg2025} provides an extensible hardware-in-the-loop environment that combines a FlatSat, ground station, orbital simulation, and interfaces for reproducing selected physical conditions. Together, these systems demonstrate the value of introducing physical hardware when electrical behavior, dynamics, interfaces, or other physical effects influence the outcome of an investigation.

MimicSat differs in its emphasis on preserving the surrounding mission while selected mission roles change realization. Hardware-based participants can be used for roles where bus behavior, device timing, firmware, finite resources, or recovery mechanisms are relevant, while software-based participants can be used where those effects are not required. A mission run may combine these participants within a mixed execution configuration while preserving the relationships among mission roles and their associated observations. This allows implementation-specific effects to be examined without simultaneously changing the mission scenario or experimental context.

\subsection{Cybersecurity environments}

Satellite cybersecurity testbeds span physical spacecraft platforms, cyber-physical laboratories, virtual ranges, network emulators, and on-orbit systems. LinkStar \cite{linkstar2022}, the Virginia Tech SmallSat cybersecurity environment \cite{vtSmallSat2020}, and related spacecraft security testbeds \cite{finke2023cscrl} support experiments involving spacecraft hardware, communications, and ground functions. The four segment Space Cybersecurity Testbed \cite{castanon2025fidelity} represents space, link, ground, and user segments and characterizes fidelity across implementation, mission behavior, data collection, threat models, attacks, and defenses. AegisSat \cite{idan2025aegissat,aegissatRepro2025} combines a physical 1U CubeSat with environmental stimulation, mission operations, attack management, and labelled telemetry collection. WALL EYE \cite{jehee2024walleye} and PwnSat \cite{pwnsat2026} provide physical platforms for security assessment and education, while the ESA Cybersecurity Laboratory \cite{esaCyberLab2026} supports hardware security, protocol fuzzing, adversary emulation, communications security, and cryptographic evaluation.

Cyber ranges and network-focused environments support repeatable security experiments involving mission and communication infrastructure. NASA IV\&V Cyber Range \cite{nasaIVVCyberRange2019} and the JPL Cyber Defense Laboratory \cite{jplCyberLab2026} support controlled studies involving mission and ground systems. MERGE/SPACE \cite{mergeSpace2024} models satellite networks with orbital events and attacker and defender environments. OpenSatRange \cite{opensatrange2026} supports satellite communication scenarios, while SOCRATE \cite{socrate2025} represents GEO links, protocol stacks, radio frequency effects, weather, and interference. StarryNet \cite{starrynet2023}, Plotinus \cite{plotinus2024}, OpenSN \cite{lu2025opensn}, and Celestial \cite{pfandzelter2022celestial} address satellite and terrestrial networking, digital twin operation, or large-scale satellite network emulation. Comparative work further shows that emulator selection can influence scalability, timing, and observed network behavior \cite{constellationEmulators2026}. The ESA Space Cyber Range \cite{esaSpaceCyberRange2026} provides another virtual environment for training and security testing.

Other platforms target narrower security objectives. SIEM4GS \cite{siem4gs2022} evaluates security monitoring in a virtual ground station environment, while a ground-based CubeSat communication testbed \cite{akpalu2026ground} examines command, telemetry, interference, and jamming using software-defined radios. QPEP \cite{qpep2023} evaluates secure communication over operational GEO links, and the SAAMD laboratory \cite{saamd2023} supports communication analysis across satellite and related cyber-physical systems. HADES \cite{hades2026} combines existing spacecraft and security software components to study telemetry and command emulation, adversary detection, and attack analysis. Space Odyssey \cite{spaceOdyssey2023} rehosts satellite firmware for security analysis. The Failure Emulator Mechanism \cite{batista2019fem} and CubeSatFI \cite{cubesatfi2021} support fault injection, while ATT\&CK SPARTA \cite{kim2026attackSparta} evaluates a ground to space attack chain. CuCD ID \cite{cucdid2026} provides labelled spacecraft cybersecurity data for intrusion detection research.

On-orbit platforms complement ground-based environments by providing access to flight hardware and operational communication paths. Moonlighter \cite{moonlighter2023} supported authorized Hack A Sat exercises, OPS SAT \cite{opsSat2024} supported in-orbit software and cybersecurity experiments, and CyberCUBE \cite{cybercube2026} supports research involving access control, jamming, spoofing, onboard monitoring, and cryptography.

Together, these environments demonstrate mature capabilities for attack evaluation, defensive testing, cyber-physical response, network experimentation, fault injection, dataset generation, and on-orbit cybersecurity research. However, MimicSat focuses on a complementary experimental problem. It preserves a common mission context while allowing selected mission roles to use different realizations and controlled interventions to be introduced during mission runs. This allows cybersecurity experiments to compare how implementation choices or intervention conditions affect subsystem behavior and mission-level consequences without requiring the surrounding mission scenario to be redefined.

\section{Discussion}
The architecture described above supports studies that require changes in implementation or operating conditions to be examined without redefining the surrounding mission. This is particularly useful when local effects must be related to subsystem interactions and mission-level consequences across repeated mission runs. The following use cases illustrate how these capabilities can support cybersecurity investigations as well as related spacecraft engineering, resilience, operations, and training studies.

\subsection{Research use cases}
\subsubsection{Cross-realization regression}
A mission role can first be assigned to a software participant and exercised through one or more mission runs under declared inputs, commands, and expected state transitions. The same role can then be assigned to an embedded participant while the surrounding mission scenario remains unchanged. An experiment can compare these runs to examine command acceptance, event ordering, latency, throughput, saturation, reset behavior, and recovery across realizations. Observed differences can then be related to modeling assumptions, interface transformations, device timing, firmware behavior, or implementation defects without simultaneously changing the surrounding mission scenario.

\subsubsection{Resource-aware resilience}
Repeated mission runs can be used to examine how controlled interventions affect individual participants and the surrounding mission. An intervention may alter a command, message, sensor input, communication condition, resource limit, or participant availability. Mission telemetry records the state reported by the affected participant, while independent observations capture communication activity, execution events, timing, and available resource behavior.
Comparing nominal and intervened mission runs makes it possible to determine whether an attack, fault, or defensive response changes subsystem behavior, delays recovery, affects resource availability, or prevents the mission from achieving its intended outcome. This separation between participant-reported state and independent observation is especially important when the affected participant may provide an incomplete account of its own behavior.

\subsubsection{Link and interface behavior}
Communication disruptions and interface changes can be varied across mission runs to examine how their effects propagate through the represented mission. Conditions such as delay, loss, contention, queue exhaustion, encoding differences, or incomplete transfers can affect command execution, payload delivery, or ground receipt. Mission telemetry remains distinct from independent bus or link observations, allowing the earliest observable divergence to be identified.
Comparing mission runs under different communication conditions can show whether an observed mission impact originates at an interface, within a participant, or later in the information path. This use case is particularly relevant when application-level success cannot be inferred from communication activity alone.

\subsubsection{Mission-impact analysis}
Mission-impact analysis examines whether a local disruption propagates beyond the affected subsystem and alters the outcome of the space mission. Power availability, communication state, payload storage, thermal conditions, and command scheduling can be varied across mission runs while preserving their relationships within the represented mission. A local reset, delayed response, rejected command, or other intervention can then be related to subsequent effects on payload acquisition, data storage, downlink, recovery, or mission completion.
Comparing these mission runs allows an experiment to distinguish a localized subsystem effect from a consequence that propagates through mission dependencies. Selected functions can also be realized differently when timing, firmware, communication, or resource behavior must be examined under the same mission scenario.

\subsubsection{Operations, replay and training}
Mission scenarios can also support operational rehearsal, training, and analysis without continuous access to the hardware mimic. Software-based mission runs can exercise schedules, communication contacts, commands, telemetry interpretation, and recovery procedures. Selected roles can then be assigned to hardware-based participants when timing, interfaces, resource behavior, or physical response are relevant to the investigation.
Completed mission runs can also be replayed for training, incident analysis, or cybersecurity detector evaluation while preserving the distinction between previously recorded observations and responses generated during replay. This allows operational procedures and recorded events to be examined repeatedly while reserving hardware participation for conditions that require physical execution.

\subsection{Limitations}
The validity of a MimicSat experiment depends on the configured participants, represented conditions, available observations, and assumptions associated with each mission run. Common mission definitions preserve functional meaning across realizations, but they do not guarantee identical timing, internal behavior, or failure responses. Similarly, an observation from one communication path does not establish the state of every participant or interaction elsewhere in the represented mission.

MimicSat prioritizes functional and informational fidelity over physical replication of spacecraft hardware. Experiments involving radiation, vacuum, structural behavior, radio frequency propagation, flight qualification, or detailed thermal conditions therefore require appropriate facilities or validated models when those effects are relevant to the research question. MimicSat instead preserves mission context, functional relationships, participant identity, state authority, and evidence provenance while allowing mission runs to use different execution configurations and selected mission roles to use different realizations. Conclusions drawn from an experiment must therefore remain within the behaviors, conditions, and observations represented by its mission runs.

\section{Conclusion}
The resulting architecture provides a common experimental basis for examining how changes at subsystem and interface levels propagate through an executing space mission. By preserving the mission definition while allowing selected roles to use different realizations, MimicSat supports comparison across software-based, hardware-based, and mixed execution configurations without conflating implementation changes with changes to the mission scenario. Source-attributable observations further allow participant-reported state and independently observed behavior to be interpreted together.
These properties support spacecraft cybersecurity studies in which attacks, faults, defensive actions, and recovery behavior must be connected to their consequences for the broader mission. They are also applicable to engineering, resilience, operations, and training studies that require controlled comparison across realizations.

\section*{Acknowledgments}

This work was supported in part by David R. Silva's Distinguished Professor Fund from the Office of the Vice President for Academic Affairs and by the Center for Space Technology and Operation Research (CSTOR) startup fund from the Office of the Vice President for Research and Innovation at The University of Texas at San Antonio.

\bibliographystyle{IEEEtran}
\bibliography{references}

@article{barcellos2023flatsat,
  author  = {Barcellos, J. C. E. and Spengler, A. W. and Seman, L. O. and {da Cruz e Silva}, R. D. and Roldan, H. P. and Bezerra, E. A.},
  title   = {FlatSat Platforms for Small Satellites: A Systematic Mapping and Classification},
  journal = {IEEE Journal on Miniaturization for Air and Space Systems},
  year    = {2023},
  volume  = {4},
  number  = {2},
  pages   = {186--198},
  doi     = {10.1109/JMASS.2023.3249044}
}

@inproceedings{finke2023cscrl,
  author    = {Finke, J. and others},
  title     = {Satellite Cybersecurity Testbed to Improve Commercial Space Security},
  booktitle = {ASCEND},
  year      = {2023},
  doi       = {10.2514/6.2023-4768}
}

@inproceedings{idan2025aegissat,
  author    = {Idan, R. and others},
  title     = {AegisSat: A Satellite Cybersecurity Testbed},
  booktitle = {NDSS Workshop on the Security of Space and Satellite Systems (SpaceSec)},
  year      = {2025},
  doi       = {10.14722/spacesec.2025.23069}
}

@inproceedings{castanon2025fidelity,
  author    = {Castanon Remy, J. L. and Chang, C. and Ear, E. and Xu, S.},
  title     = {Space Cybersecurity Testbed: Fidelity Framework, Example Implementation, and Characterization},
  booktitle = {NDSS Workshop on the Security of Space and Satellite Systems (SpaceSec)},
  year      = {2025},
  doi       = {10.14722/spacesec.2025.23042}
}

@techreport{nasa7009,
  author = {{National Aeronautics and Space Administration}},
  title = {Standard for Models and Simulations},
  institution = {NASA},
  number = {NASA-STD-7009B},
  year = {2024},
  url = {https://standards.nasa.gov/standard/NASA/NASA-STD-7009}
}

@inproceedings{grubb2016,
  author = {Matthew Grubb and Justin Morris and Scott Zemerick and John Lucas},
  title = {{NASA} Operational Simulator for Small Satellites ({NOS3}): Tools for Software-based Validation and Verification of Small Satellites},
  booktitle = {30th Annual AIAA/USU Conference on Small Satellites},
  year = {2016},
  url = {https://digitalcommons.usu.edu/smallsat/2016/S2CDH/1/}
}

@inproceedings{lucas2023,
  author = {John P. Lucas and Matthew D. Grubb and Justin R. Morris and Mark D. Suder and Scott A. Zemerick},
  title = {{NASA} Operational Simulator for {SmallSats} ({NOS3}) -- Design Reference Mission},
  booktitle = {37th Annual Small Satellite Conference},
  year = {2023},
  note = {Paper SSC23-XIII-06},
  url = {https://ntrs.nasa.gov/citations/20230010613}
}

@article{kiesbye2019,
  author = {Jonis Kiesbye and David Messmann and Maximilian Preisinger and Gonzalo Reina and Daniel Nagy and Florian Schummer and Martin Mostad and Tejas Kale and Martin Langer},
  title = {Hardware-In-The-Loop and Software-In-The-Loop Testing of the {MOVE-II CubeSat}},
  journal = {Aerospace},
  volume = {6},
  number = {12},
  pages = {130},
  year = {2019},
  doi = {10.3390/aerospace6120130},
  url = {https://doi.org/10.3390/aerospace6120130}
}

@misc{nos3validation,
  author = {{NASA Independent Verification and Validation Facility}},
  title = {{STF} Verification and Validation ({V\&V}) Plan},
  howpublished = {NOS3 documentation, version 1.07.03},
  note = {Accessed September 10, 2026},
  url = {https://nos3.readthedocs.io/en/v1_07_03/STF_VV_Plan.html}
}

@inproceedings{greenberg2025,
  author = {Andrew Greenberg and Glenn LeBrasseur and Max Eltzroth and Rose Edington and Anna Meneely and Maxwell Millar},
  title = {An Inexpensive, Open, and Extensible Hardware-in-the-Loop {CubeSat} Test and Simulation Infrastructure},
  booktitle = {39th Annual Small Satellite Conference},
  year = {2025},
  url = {https://digitalcommons.usu.edu/smallsat/2025/all2025/196/}
}

@misc{esa2021,
  author = {{European Space Agency}},
  title = {Opened-out `{FlatSat}' for {CubeSat} Testing},
  year = {2021},
  month = oct,
  howpublished = {\href{https://www.esa.int/Enabling_Support/Space_Engineering_Technology/Opened-out_FlatSat_for_CubeSat_testing}{ESA Space Engineering and Technology}}
}

@article{sternberg2018,
  title={JPL Small Satellite Dynamics Testbed Simulation: Validated Models for Predicting On-Orbit Performance},
  author={Sternberg, D and Pong, C and Filipe, N and Mohan, S and Johnson, S and Jones-Wilson, L},
  journal={J. Spacecraft and Rockets},
  volume={55},
  pages={322--334},
  year={2018}
}

@inproceedings{rivera2023,
  title={Multi-platform small satellite dynamics testbed},
  author={Rivera, Kalani Danas and Sternberg, David and Lo, Kevin and Mohan, Swati},
  booktitle={2023 IEEE Aerospace Conference},
  pages={1--7},
  year={2023},
  organization={IEEE}
}

@misc{moonlighter2023,
  author = {{The Aerospace Corporation}},
  title  = {Moonlighter Illuminates a Path to Resiliency},
  year   = {2023},
  url    = {https://aerospace.org/article/aerospaces-moonlighter-illuminates-path-resiliency}
}

@inproceedings{siem4gs2022,
  author    = {Law, Yee Wei and Slay, Jill},
  title     = {{SIEM4GS}: Security Information and Event Management for a Virtual Ground Station Testbed},
  booktitle = {European Conference on Cyber Warfare and Security},
  year      = {2022},
  url       = {https://papers.academic-conferences.org/index.php/eccws/article/view/228}
}

@inproceedings{hades2026,
  author    = {Chan, Arnold and Seo, Sam and James, Ady},
  title     = {Defending the Cosmos: Honeypot-based Adversary Detection and Emulation System},
  booktitle = {25th European Conference on Cyber Warfare and Security},
  year      = {2026},
  doi       = {10.34190/eccws.25.1.4647}
}

@techreport{akpalu2026ground,
  author      = {Akpalu, Ruth},
  title       = {Evaluating a Ground-Based Testbed for CubeSat Communication},
  institution = {Old Dominion University},
  year        = {2026},
  url         = {https://vsgc.odu.edu/wp-content/uploads/2026/04/Akpalu_Ruth_VSGC_Final_Paper_Submission_2026.pdf}
}

@misc{vtSmallSat2020,
  author = {{Commonwealth Cyber Initiative and Virginia Tech}},
  title  = {SmallSat Cybersecurity and Resiliency},
  year   = {2020},
  url    = {https://cyberinitiative.org/research/funded-projects/2020/2020-cybersecurity-research-collaboration/SmallSat-cybersecurity-and-resiliency.html}
}

@mastersthesis{jehee2024walleye,
  author = {Jehee, Wouter},
  title  = {{WALL-EYE}: Taking a Look at {CubeSat} Security},
  school = {Delft University of Technology},
  year   = {2024},
  url    = {https://repository.tudelft.nl/file/File_3e1d9b8b-ffc8-47da-bca1-0925fa9eaf4b}
}

@misc{pwnsat2026,
  author = {{PwnSat Project}},
  title  = {{PwnSat}: Vulnerable-by-Design Satellite Hardware and Mission Platform},
  year   = {2026},
  url    = {https://github.com/ElectronicCats/PWNCUBE},
  note   = {Accessed August 2026}
}

@inproceedings{linkstar2022,
  author    = {Santangelo, Andrew D. and Falco, Gregory and Viswanathan, Arun},
  title     = {The {LinkStar} Cybersecurity Sandbox, a Platform to Test Small Satellite Vulnerabilities within the Community: Updates and Lessons Learned},
  booktitle = {AIAA SCITECH 2022 Forum},
  year      = {2022},
  doi       = {10.2514/6.2022-0239}
}

@techreport{nasaIVVCyberRange2019,
  author      = {Bailey, Brandon},
  title       = {{NASA IV\&V's Cyber Range for Space Systems}},
  institution = {NASA Independent Verification and Validation Program},
  year        = {2019},
  number      = {GSFC-E-DAA-TN65725},
  url         = {https://ntrs.nasa.gov/citations/20190001085}
}

@misc{jplCyberLab2026,
  author = {{NASA Jet Propulsion Laboratory}},
  title  = {Cyber Defense Laboratory},
  year   = {2026},
  url    = {https://www.jpl.nasa.gov/site/research/research-community/laboratories-facilities/cyber-defense-laboratory/},
  note   = {Accessed August 2026}
}

@inproceedings{mergeSpace2024,
  author    = {Collins, M. Patrick and Hussain, Alefiya and Walters, J. P. and Ardi, Calvin and Tran, Chris and Schwab, Stephen},
  title     = {{Merge/Space}: A Security Testbed for Satellite Systems},
  booktitle = {Workshop on the Security of Space and Satellite Systems},
  year      = {2024},
  doi       = {10.14722/spacesec.2024.23074},
  url       = {https://www.ndss-symposium.org/wp-content/uploads/spacesec2024-74-paper.pdf}
}

@inproceedings{saamd2023,
  author     = {Costin, Andrei and Turtiainen, Hannu and Khandker, Syed and Hamalainen, Timo},
  title      = {Towards a Unified Cybersecurity Testing Lab for Satellite, Aerospace, Avionics, Maritime, Drone Technologies and Communications},
  booktitle  = {Workshop on the Security of Space and Satellite Systems},
  year       = {2023},
  doi        = {10.14722/spacesec.2023.239280},
  eprint     = {2302.08359},
  eprinttype = {arXiv}
}

@inproceedings{qpep2023,
  author    = {Huwyler, Julian and Pavur, James and Tresoldi, Giorgio and Strohmeier, Martin},
  title     = {{QPEP} in the Real World: A Testbed for Secure Satellite Communication Performance},
  booktitle = {Workshop on the Security of Space and Satellite Systems},
  year      = {2023},
  doi       = {10.14722/spacesec.2023.239792}
}

@misc{opensatrange2026,
  author = {{OpenSatRange Project}},
  title  = {{OpenSatRange}: Open Cyber Range for Satellite Communication Scenarios},
  year   = {2026},
  url    = {https://romars.tech/en/project/opensatrange/},
  note   = {Completed platform and validated training scenarios, accessed August 2026}
}

@inproceedings{socrate2025,
  author    = {Santorsola, Alessandro and Mammone, Daniele and Longari, Stefano and Topputo, Francesco and Merg{\`e}, Matteo},
  title     = {An End-to-End {GEO} Satellite Links Simulation Framework for Cyber Range Applications},
  booktitle = {3S Symposium},
  year      = {2025},
  url       = {https://indico.esa.int/event/571/attachments/7211/13612/An%20End-to-End%20GEO%20Satellite%20Links%20Simulation.pdf}
}

@misc{esaCyberLab2026,
  author = {{European Space Agency}},
  title  = {Cybersecurity Laboratory},
  year   = {2026},
  url    = {https://technology.esa.int/lab/cybersecurity-laboratory},
  note   = {Accessed August 2026}
}

@misc{esaSpaceCyberRange2026,
  author = {{European Space Agency}},
  title  = {Space Cyber Range},
  year   = {2026},
  url    = {https://resilience.esa.int/archives/projects/space-cyber-range-scr},
  note   = {Current project description, accessed August 2026}
}

@inproceedings{aegissatRepro2025,
  author    = {Peled, Roy and Idan, Roee and Cohen Galor, Tomer and Chodeda, Ofir and Marcus, Eli and Shabtai, Asaf and Elovici, Yuval},
  title     = {A Reproducible and Open-Source Testbed for Satellite Cybersecurity},
  booktitle = {ACM Conference on Reproducibility and Replicability},
  pages     = {51--61},
  year      = {2025},
  doi       = {10.1145/3736731.3746144}
}

@article{kim2026attackSparta,
  author  = {Kim, Kyeong-Ho and Ryou, Jae-Cheol},
  title   = {Integrated {ATT\&CK--SPARTA} Threat Modeling and Security Validation for {NewSpace} Systems: An Emulation-Based Testbed Approach},
  journal = {IEEE Access},
  year    = {2026},
  doi     = {10.1109/ACCESS.2026.3707097}
}

@article{cucdid2026,
  author  = {Fayyaz, Yasamin and Yang, Li and El-Khatib, Khalil},
  title   = {{CubeSat} Cybersecurity Dataset for Intrusion Detection ({CuCD-ID}): Labelled {NOS3/cFS} Telemetry with {COSMOS} Reproduction Scripts},
  journal = {Data in Brief},
  volume  = {65},
  pages   = {112598},
  year    = {2026},
  doi     = {10.1016/j.dib.2026.112598}
}

@inproceedings{cubesatfi2021,
  author    = {Paiva, David and Duarte, Jose Marcelo Lima and Lima, Raffael and Carvalho, Manoel and Mattiello-Francisco, Fatima and Madeira, Henrique},
  title     = {{CubeSatFI}: Fault Injection Platform for Affordable Verification and Validation of {CubeSats} Software},
  booktitle = {Latin-American Symposium on Dependable Computing},
  year      = {2021},
  doi       = {10.1109/LADC53747.2021.9672584}
}

@article{batista2019fem,
  author  = {Batista, Carlos Leandro Gomes and Weller, Anderson Coelho and Martins, Eliane and Mattiello-Francisco, Fatima},
  title   = {Towards Increasing Nanosatellite Subsystem Robustness},
  journal = {Acta Astronautica},
  volume  = {156},
  pages   = {187--196},
  year    = {2019},
  doi     = {10.1016/j.actaastro.2018.11.011}
}

@inproceedings{spaceOdyssey2023,
  author    = {Willbold, Johannes and Schloegel, Moritz and Vogele, Manuel and Gerhardt, Maximilian and Holz, Thorsten and Abbasi, Ali},
  title     = {Space Odyssey: An Experimental Software Security Analysis of Satellites},
  booktitle = {2023 IEEE Symposium on Security and Privacy},
  year      = {2023},
  url       = {https://jwillbold.com/paper/willbold2023spaceodyssey.pdf}
}

@misc{opsSat2024,
  author = {{European Space Agency}},
  title  = {{OPS-SAT} Mission Completion},
  year   = {2024},
  url    = {https://www.esa.int/Enabling_Support/Operations/OPS-SAT_Space_Lab_closes},
  note   = {Mission ended in May 2024}
}

@misc{cybercube2026,
  author = {{GMV}},
  title  = {{CyberCUBE}, ESA's Mission to Strengthen the Cybersecurity of Space Systems, Successfully Launched from California},
  year   = {2026},
  url    = {https://www.gmv.com/en/communication/press-room/press-releases/space-safety/cybercube-mission-strengthen-cybersecurity},
  note   = {Accessed August 2026}
}

@inproceedings{starrynet2023,
  author    = {Lai, Zeqi and Li, Hewu and Wu, Jihao and Zhang, Qian and Li, Ye and Li, Yiming and Cui, Junchen and Wu, Jilong},
  title     = {{StarryNet}: Empowering Researchers to Evaluate Futuristic Integrated Space and Terrestrial Networks},
  booktitle = {20th USENIX Symposium on Networked Systems Design and Implementation},
  year      = {2023},
  url       = {https://www.usenix.org/conference/nsdi23/presentation/lai}
}

@misc{plotinus2024,
  author     = {Gao, Yue and Qiu, Kun and Chen, Zhe and Zhu, Wenjun and Zhang, Qi and Luo, Handong and Lin, Quanwei and Yang, Ziheng and Liu, Wenhao},
  title      = {{Plotinus}: A Satellite Internet Digital Twin System},
  year       = {2024},
  eprint     = {2403.08515},
  eprinttype = {arXiv}
}

@misc{constellationEmulators2026,
  author     = {Cionca, Victor and Szabo, Ferenc and Vasilev, Stanimir and Smyth, Dylan},
  title      = {An Experimental Evaluation of Satellite Constellation Emulators},
  year       = {2026},
  eprint     = {2604.04498},
  eprinttype = {arXiv},
  note       = {Preprint}
}

@inproceedings{pfandzelter2022celestial,
  title={Celestial: Virtual software system testbeds for the leo edge},
  author={Pfandzelter, Tobias and Bermbach, David},
  booktitle={Proceedings of the 23rd ACM/IFIP International Middleware Conference},
  pages={69--81},
  year={2022}
}

@article{lu2025opensn,
  title     = {OpenSN: An Open Source Library for Emulating LEO Satellite Networks},
  author    = {Lu, Wenhao and Wang, Zhiyuan and Zhang, Hefan and Zhang, Shan and Luo, Hongbin},
  journal   = {IEEE Transactions on Parallel and Distributed Systems},
  year      = {2025},
  month     = jun,
  volume    = {36},
  number    = {8},
  pages     = {1574--1590},
  publisher = {Institute of Electrical and Electronics Engineers (IEEE)},
  doi       = {10.1109/TPDS.2025.3575920},
  url       = {https://arxiv.org/pdf/2507.03248}
}

\end{document}